\documentclass[prx,aps,epsf,showpacs,superscriptaddress,longbibliography,nofootinbib,twocolumn,tightenlines]{revtex4-1}
\usepackage{graphicx} 
\usepackage{braket}
\usepackage{quantikz}
\usepackage{multirow}
\usepackage{comment}
\usepackage{url}
\newcommand{\drew}[1]{\textcolor{orange}{\it 
 Drew: #1}}

\begin{document}
\title{
Breaking even with magic: demonstration of a high-fidelity logical non-Clifford gate}
\author{Shival Dasu} 
\affiliation{Quantinuum, 303 S. Technology Ct., Broomfield, Colorado 80021, USA}

\author{Simon Burton} 
\affiliation{Quantinuum, Partnership House, Carlisle Place, London SW1P 1BX, UK}

\author{Karl Mayer}
\affiliation{Quantinuum, 303 S. Technology Ct., Broomfield, Colorado 80021, USA}

\author{David Amaro}
\affiliation{Quantinuum, Partnership House, Carlisle Place, London SW1P 1BX, UK}

\author{Justin A. Gerber}
\affiliation{Quantinuum, 303 S. Technology Ct., Broomfield, Colorado 80021, USA}

\author{Kevin Gilmore} 
\affiliation{Quantinuum, 303 S. Technology Ct., Broomfield, Colorado 80021, USA}

\author{Dan Gresh}
\affiliation{Quantinuum, 303 S. Technology Ct., Broomfield, Colorado 80021, USA}

\author{Davide DelVento}
\affiliation{Quantinuum, 303 S. Technology Ct., Broomfield, Colorado 80021, USA}

\author{Andrew C. Potter}
\affiliation{Quantinuum, 303 S. Technology Ct., Broomfield, Colorado 80021, USA}

\author{David Hayes}
\affiliation{Quantinuum, 303 S. Technology Ct., Broomfield, Colorado 80021, USA}

\date{February 2025}

\begin{abstract}
    Encoding quantum information to protect it from errors is essential for performing large-scale quantum computations. Performing a universal set of quantum gates on encoded states demands a potentially large resource overhead and minimizing this overhead is  key for the practical development of large-scale fault-tolerant quantum computers. We propose and experimentally implement a magic-state preparation protocol to fault-tolerantly prepare a pair of logical magic states in a [[6,2,2]] quantum error-detecting code using only eight physical qubits. Implementing this protocol on H1-1, a 20 qubit trapped-ion quantum processor, we prepare magic states with experimental infidelity $7^{+3}_{-1}\times 10^{-5}$ with a $14.8^{+1}_{-1}\%$ discard rate and use these to perform a fault-tolerant non-Clifford gate, the controlled-Hadamard (CH), with logical infidelity $\leq 2.3^{+9}_{-9}\times 10^{-4}$. Notably, this significantly outperforms the unencoded physical CH infidelity of $10^{-3}$. 
    Through circuit-level stabilizer simulations, we show that this protocol can be self-concatenated to produce extremely high-fidelity magic states with low space-time overhead in a [[36,4,4]] quantum error correcting code, with logical error rates of $6\times 10^{-10}$ ($5\times 10^{-14}$) at two-qubit error rate of $10^{-3}$ ($10^{-4}$) respectively.
\end{abstract}
\maketitle

\begin{figure*}[t]
    \centering
    \includegraphics[width=1.0\linewidth]{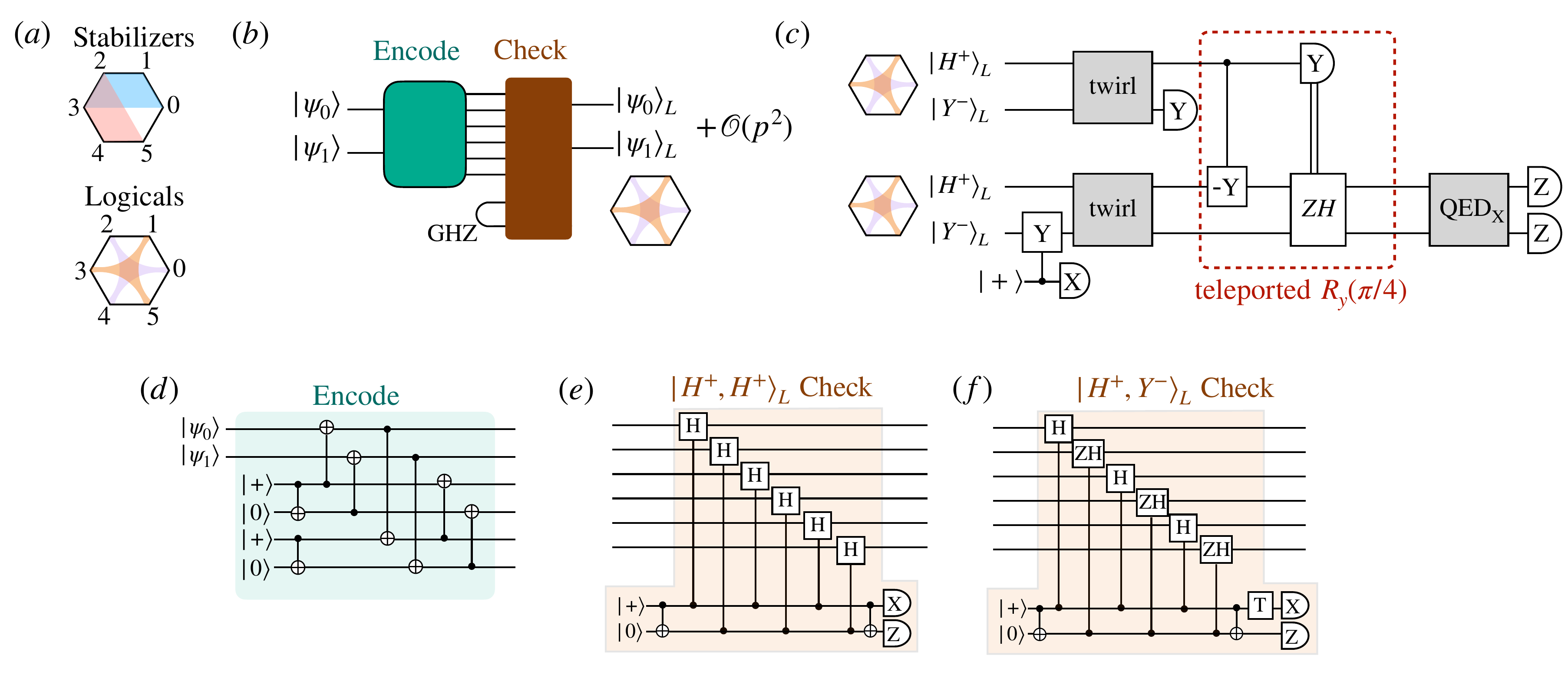}
    \caption{{\bf Preparing and benchmarking magic states:} (a) Schematic of the the self-dual $H_6$ [[6,2,2]] code with physical qubits depicted as vertices of a hexagon, with generating stabilizers and logical operators indicated by shaded shapes. (b) Schematic of state preparation protocol where an arbitrary state encoder encodes physical states $\ket{\psi_{0,1}}$ into the logical states of a $H_6$ code (circuit shown in (d)), followed by a check stage using an ancilla (physical) Bell/GHZ pair. Although the encoder circuit works for arbitrary inputs, the check stage requires that $\ket{\psi_{0,1}}$ be stabilized by transversal gates. Specific checks with for preparing two or one magic states for $\ket{\psi_{0,1}} \in \{\ket{H^+},\ket{Y^-}\}$ are shown in (e,f) respectively.
    As explained in the text, this protocol may be run on physical qubits, in which case the output is two ``level-1" magic states in the [[6,2,2]] code, or it may be run one concatenation level higher, in which case the two wires represent logical qubits in two separate [[6,2,2]] blocks and the CNOTs in the encoding circuit are transversal and the output is four ``level-2" magic states. 
    (c) A fault-tolerant circuit for benchmarking the combined infidelity of two logical magic states on different code blocks. 
    Twirling stages, described in the text, are used to decohere errors to obtain rigorous fidelity bounds.
    In these experiments, data is post-selected on having a $+1$ measurement outcome for the $\bar{Y}_2$ measurements on the second qubit of each block, as well as each of the x-syndrome measurements ($QED_X$ block). For preparing resource states offline in a fault tolerant quantum computing architecture, this post-selection would be replaced by a repeat-until-success protocol with modest additional spatial overhead.
    }
    \label{fig:magicprepcombined}
\end{figure*}


Quantum error correction (QEC) allows quantum computers to exponentially suppress errors, at the cost of encoding information redundantly in a quantum error correcting code. Once the information is protected by a QEC code, it must be manipulated fault-tolerantly, meaning, in a way that does not spread errors uncontrollably. For a given QEC code, the fault-tolerant operations which are possible are highly constrained \cite{Eastin_2009}, and devising a universal set of fault-tolerant routines is a non-trivial task. Often the most challenging routine is performing a non-Clifford gate, which is a kind of quantum operation which is necessary in order to perform quantum computations which cannot be simulated classically and are, therefore, key to achieving any quantum advantage. These operations are often implemented fault-tolerantly by consuming special resource states known as ``magic states," which themselves are either prepared fault-tolerantly or prepared noisily and then distilled~\cite{Bravyi2005}. Implementing a comprehensive set of fault-tolerant routines has arguably been studied most thoroughly for the surface code \cite{Fowler2012, Gidney_2019, gidney2024}, which possesses a local connectivity well-suited to planar architectures.

For architectures with all-to-all connectivity between qubits such as the QCCD architecture\cite{Wineland98}, codes that are non-local may significantly reduce the overheads of QEC due to their higher encoding rates and distance. 
Although many promising high-rate, high-distance QEC codes have been discovered~\cite{panteleev2022, leverrier2022}, efficiently manipulating the quantum information in these codes is still an open area of research. One strategy is to start with a small, high-rate code and concatenate it with itself~\cite{Goto_2024} to retain a reasonably high encoding rate and implement logical gates by performing gates which are native to lower levels of concatenation.  While these concatenated routines are efficient, the overhead of magic state distillation~\cite{Bravyi2005,Bravyi2012} for such codes is high relative to the state-of-the-art in magic state preparation for the surface and color code \cite{Fowler2012,Gidney_2019,Chamberland_2020,gidney2019flexible,gidney2024,lee2025},  such as magic state cultivation \cite{gidney2024}. 
Rather than distilling a higher-quality magic state from multiple noisy copies~\cite{Bravyi2005}, the latter schemes instead prepare a (possibly faulty) logical $\ket{H^+}=\cos\frac\pi 8 \ket{0} + \sin\frac\pi 8 \ket{1}$ magic state ($+1$ eigenstate of the Hadamard, $H$, operator), measure the logical $H$ operator of the code using ancilla qubits to verify this state, and abort and retry the protocol if the check is unsuccessful. 
This technique is sometimes referred to as ``0-level distillation"~\cite{hirano2024, itogawa2024efficient}.

In this work, we adapt the 0-level distillation ideas to an $H_6$ code~\cite{Knill_2005, Jones2013} with parameters [[6,2,2]], to prepare two $\ket{H^+}$ magic states with a logical failure rate $O(p^2)$, which is suppressed relative to the physical gate failure rate, $p$. This error scaling implies that these magic states can be fault-tolerantly used in a d=4 error-correcting code, although, due to limited qubit resources, we benchmark them in the [[6,2,2]] code on H1-1, a 20 qubit, trapped-ion quantum processor made by Quantinuum. We show how to fault-tolerantly send these magic states to the $[[16,4,4]]$ code in the Discussion section and Appendix \ref{app: code switch}.

Including two ancillae to perform the $H$-check, this protocol requires only eight physical qubits to prepare two magic states. 
This protocol is a refinement of \cite{Goto_magic} and \cite{Jones2013}, but achieves the $O(p^2)$ failure rate at one lower level of concatenation than \cite{Goto_magic}, significantly reducing the required number of qubits and circuit complexity, while maintaining robustness to noisy two-qubit gates in the state-encoding and fault-checking circuitry, which were assumed to be noiseless in \cite{Jones2013}.
We experimentally benchmark this eight qubit magic state preparation protocol on Quantinuum's system model H1 trapped-ion quantum processor, and use it to fault-tolerantly implement and benchmark a logical non-Clifford gate, a controlled-Hadamard (CH), with significantly lower error than the physical CH gate.

This small, distance-two protocol can be self-concatenated to achieve further error suppression.
We demonstrate that self-concatenating the distance-two protocol can produce four magic states in a [[36,4,4]] code with failure rate $O(p^4)$, using approximately forty physical qubits per magic state, and running in depth 15. This is similar to the concatenation in \cite{Jones2013}, but uses flagged, encoded ancillae to additionally protect against gate and measurement errors, as in \cite{Goto_magic}, but again with one less layer of encoding.
Just as in magic state cultivation~\cite{gidney2024}, before use in a general fault-tolerant computation, the high-fidelity magic states produced by this protocol require an ``escape" stage which sends the magic states to a higher-distance QEC code while maintaining their fault distance. This escape stage can be efficiently accomplished with level-raising teleportation~\cite{Goto_magic} to fault-tolerantly send the magic states into codes with higher levels of concatenation.

The paper is structured as follows. First we introduce an eight-qubit magic state preparation protocol for a [[6,2,2]] quantum error-detecting code and discuss its fault-tolerance. 
Then we describe our benchmarking procedure and a variant of the protocol that fault-tolerantly produces a single magic state with properties that enable rigorous fidelity arguments. 
We measure the state fidelity of the magic states produced, as well as the fidelity of a fault-tolerant logical $R_y(\pi/4)$ rotation and a fault-tolerant controlled Hadamard. From our benchmarking procedure, we estimate that our protocol has a discard rate of $14.8^{+1}_{-1}\%$ and produces magic states of infidelity $7^{+3}_{-1}\times10^{-5}$ on H1-1. This is perhaps the lowest magic state infidelity ever experimentally demonstrated on any platform while fault-tolerantly using a code of non-trivial distance \cite{Lacroix:2024vls, Rodriguez:2024bhh, Pogorelov:2024zvv, Kim:2024vmw,Ye:2023hxg,Gupta:2023zei, Anderson2021,Postler:2021ddz, Marques:2021kev, Egan:2020kdu}. Although our results produce a magic state in a $d=2$ code, in actual computations, these magic states will be sent via level-raising teleportation and code switching to a code of distance 4, the [[16,4,4]] code. For more details, see the Discussion section and Appendix \ref{app: code switch}. Over the course of our experiments, another set of experiments were carried out on the Quantinuum H-series hardware which achieved perhaps the lowest magic state infidelity in a code with distance $\geq3$, see~\cite{Daguerre:2025} for a detailed write-up of those experiments which includes a summary table of other similar experimental demonstrations. While the fidelity of the logical magic state and the logical $R_y(\pi/4)$ rotation are both less than $10^{-4}$, we still do not break even on these single-qubit (1Q) gates because the state prep fidelity and 1Q gate infidelity of H1 is extremely small, $\sim 10^{-5}$. However, we break even on a non-Clifford two qubit gate using these magic states, since the average error of the fault-tolerant implementation is at most $2.3^{+9}_{-9}\times 10^{-4}$ compared to the un-encoded (``physical") version of this gate, which has infidelity at least $1.0^{+1}_{-1}\times 10^{-3}$. Finally, we simulate and discuss the concatenated protocol.

\section{Eight Qubit Magic State Preparation}
The eight-qubit magic state preparation protocol is based on the $H_6$ code = with parameters [[6,2,2]] which we present with a generating set of stabilizers: $XXXXII,IIXXXX, ZZZZII,IIZZZZ$
and logical operators: $\bar{X}_0 = XIXIXI, \bar{X}_1 = IXIXIX, \bar{Z}_0 = ZIZIZI, \bar{Z}_1 = IZIZIZ$ (shown graphically in Fig.~\ref{fig:magicprepcombined}a).
In this presentation, the code is clearly self-dual under $X\leftrightarrow Z$, and a transversal physical Hadamard induces a logical Hadamard on both logical qubits: $H^{\otimes 6} = \bar{H}_0\bar{H}_1.$ For this reason, the code was dubbed the $H_6$ code in \cite{Jones2013}, although it was originally introduced as the $C_6$ code in \cite{Knill_2005}.

Fig.~\ref{fig:magicprepcombined}b shows a protocol to prepare a pair of magic states, $\ket{H^+,H^+}_L$, where $\ket{H^+} = \cos(\pi/8)\ket{0} + \sin(\pi/8)\ket{1}$ is the $+1$ eigenstate of the Hadamard operator, and the $L$ subscript denotes the logically-encoded state. 
The protocol begins by using an arbitrary-state encoder circuit~\cite{Goto_magic} (Fig.~\ref{fig:magicprepcombined}d) to encode two magic states in the $[[6,2,2]]$ code. 
Any single gate failure in the encoder circuit can cause a logical error in at most one of the two logical qubits.
The occurrence of such a single-logical error can be detected by measuring $\bar{H}_0\bar{H}_1$ via controlled-Hadamard gates from a physical ancilla as shown in Fig.~\ref{fig:magicprepcombined}e. An additional flag qubit detects whether $X$ errors on the ancilla spread to higher weight errors on the code block.
For measurement outcome $\bar{H}_0\bar{H}_1 = +1$, the logical state is in the span of $\ket{H^+,H^+}_L\>$ and $ \ket{H^-,H^-}_L\>$, where $\ket{H^-} = Y\ket{H^+}$ is the $-1$ eigenvector of $H$ and $Y$ is the Pauli-$Y$ operator. The $\ket{H^-,H^-}_L$ component results only from logical errors on both qubits, which require at least two gate faults, and occurs with probability $O(p^2)$. An error occurring with probability $O(p)$ that is not a logical error can be detected by subsequent syndrome extraction.

\section{Experimental Results}

\subsection{Benchmarking the Magic-State preparation}

While the magic-state preparation protocol is capable of producing two magic states per codeblock, for reasons explained in Appendix~\ref{app: MSbenchmarkingdetails} it is simpler and more reliable to bound the fidelity by isolating a single magic state. 
To this end, we modify the magic state preparation procedure to prepare a logical state $\ket{H^+,Y^-}_L$, where $\ket{Y^-} = \frac{1}{\sqrt{2}}(\ket{0} - i\ket{1})$ is the -1 eigenstate of the $Y$ operator, as shown in Figure \ref{fig:magicprepcombined}f. 
This modification uses a nearly identical circuit, remains fault-tolerant, and does not significantly alter the fidelity of the $\ket{H^+}$ state in the first logical qubit. As explained in Appendix~\ref{app: MSbenchmarkingdetails}, the choice of the $Y^-$ basis state for the second logical qubit enables a convenient twirling operation to break the coherence of errors and obtain rigorous fidelity bounds. We call this protocol the experimental protocol, as it is the one used in all logical benchmarking experiments on H1-1.

A conceptually straightforward way to estimate the fidelity of the magic states produced by this protocol would be to measure them in the $X$, $Y$, and $Z$ bases, compute expectation values to estimate the logical state, and directly compute the fidelity. However, since the expected infidelity of these magic states is $\bar{\mathcal{F}}_M\lesssim 10^{-4}$, obtaining an accurate statistical estimate in this fashion would require a prohibitively large number of shots $\mathcal{O}\left(\bar{\mathcal{F}}_M^{-2}\right) \sim 10^{8}.$  Instead, we prepare magic states in two different code blocks using this protocol, and use one to rotate the other back into the computational basis by a $R_y(-\pi/4)$ rotation. The circuit for performing $R_y(\pm \pi/4)$ rotations using an $\ket{H}$ state is shown in Fig.~\ref{fig:magicprepcombined}c. 
An ideal execution of this protocol deterministically yields $\ket{0}_L$, and the empirical probability of obtaining the ideal outcome allows for an estimate of fidelity of the magic state with only $\mathcal{O}(\bar{\mathcal{F}}_M^{-1})\sim 10^4$ shots. 

Since the benchmarking protocol involves two magic states, its failure rate upper-bounds twice the infidelity of each magic state (see Appendix A for a formal derivation).
Implementing this protocol on the H1 quantum processor with 20 trapped-ion qubits, using 120,000 shots reveals a magic state infidelity of:
\begin{align}
    \bar{\mathcal{F}}_M \leq 7_{-1}^{+3}\times 10^{-5}.
\end{align} 
Shots that either fail to pass the $H$-check or yield a non-trivial $X$-syndrome in the $\rm QED_X$ stage are rejected, resulting in an acceptance rate of $85.2^{+1}_{-1}\%$.
To assess the importance of these error-detection gadgets, we compare this fault-tolerant benchmarking circuit to a non-fault-tolerant version that omits the $H$-check and $X$-syndrome extraction step, while still using the syndrome information inferred from measuring out the two codeblocks, and observe an order of magnitude increase in the infidelity, $\bar{\mathcal{F}}_{M,\rm non-FT} \leq (7_{-1}^{+1})\times 10^{-4}$. We approximate twirling by submitting $30,000$ shots for each of the four possibilities where the $\bar{X}_2\bar{H}_1\bar{H}_2$ operator either is or is not applied for the fault-tolerant version. For the non-fault-tolerant version, since the error rate is an order of magnitude higher, we submit $10,000$ shots for each possibility to obtain adequate bounds on the uncertainty.

\subsection{Benchmarking the $R_y(\pi/4)$ Rotation}

\begin{figure}
    \centering
    \includegraphics[width=0.5\textwidth]{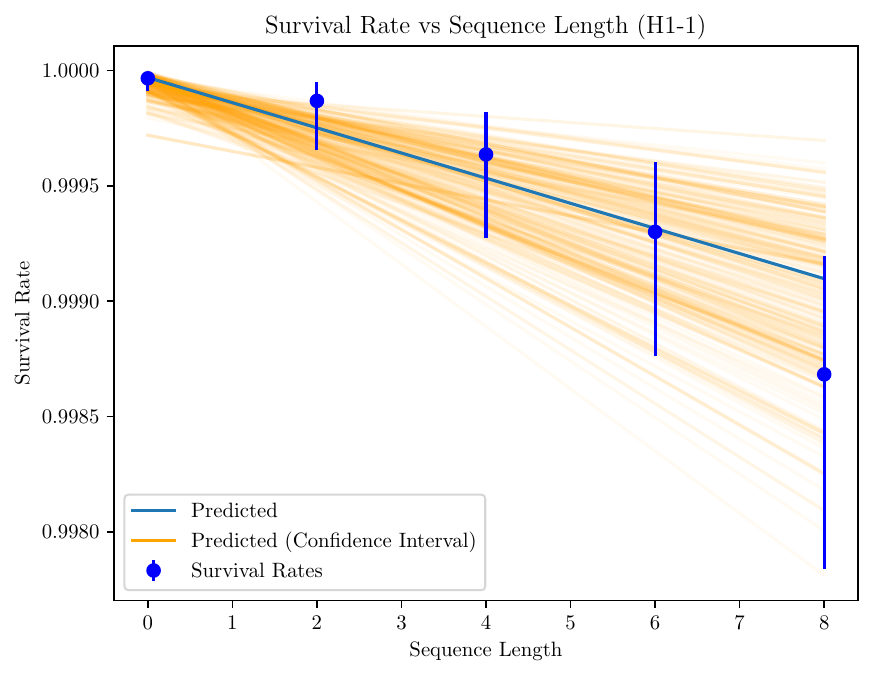}
    \caption{On H1-1, a variable number of logical $R_y(-\pi/4)$ rotations were performed on one of the logical qubits in a [[6,2,2]] code using magic states generated by our experimental protocol. The blue line is the model under the maximum likelihood estimates (MLE) for the model parameters given the likelihood function that arises from the observed data. Each orange line is a random sample from a Markov chain Monte Carlo (MCMC) sampling of the model parameters under the same likelihood function. The spread in the MCMC samples lines gives a sense for the variance in the model predictions given the data. Using MLE to estimate the model parameters yields an infidelity of $(7^{+3}_{-2})\times10^{-5}$ for a logical $R_y(-\pi/4)_L$ rotation, where the confidence interval is derived from MCMC.}
\label{fig:RyPi4RamseyPlot}
\end{figure}

In addition to benchmarking the magic-state fidelity, we also benchmark the fidelity of the fault-tolerant logical $R_y(-\pi/4)$ rotation in Fig.~\ref{fig:magicprepcombined}c by applying a sequence of an increasing number of $Y$-rotations similar to Ref.~\cite{mayer2024}. 
An important component for obtaining fidelity bounds is to introduce a twirling step to tailor the logical error channel into a simple form.
If the state is $\ket{H^+,Y^-}_L$ and we twirl by $\bar{X}_2\bar{H}_1\bar{H}_2$ before gate teleportation, the reasoning in \cite{Temme_2017} shows that the noisy implementation of the $R_y(-\pi/4)$ gate will be 
\begin{equation}\label{eq:yflip channel}
    \tilde{\mathcal{R}}_y = (1-\epsilon)\mathcal{R}_y + \epsilon \mathcal{Y} \circ \mathcal{R}_y 
\end{equation}
where $\tilde{\mathcal{R}}_y$ is the channel for the noisy $R_y(-\pi/4)$ gate, $\mathcal{R}_y$ is the channel for the ideal gate, $\mathcal{Y}$ is a $Y$-flip channel and $\epsilon$ is the infidelity of the magic state used for the gate teleportation. The average infidelity of the logical $R_y(-\pi/4)$ rotation is given by $\frac{2}{3}\epsilon.$

We prepare an initial $\ket{0}_L$ state and apply the $R_y(-\pi/4)$ gate $L$ times with $L \in \{0,2,4,6,8\}$, sending the state from $\ket{0}_L$ to $\ket{-}_L, \ket{1}_L,\ket{+}_L$ and finally back to $\ket{0}_L$. Before each gate teleportation, we twirl the magic state by applying the $\bar{X}_2\bar{H}_1\bar{H}_2$ operator conditioned on a randomly generated bit. The probability of obtaining the expected state will decay exponentially in $L$ and is fit to the model:
\begin{equation}\label{eq: Ramsey Curve}
    P(L) = \frac{1}{2} + \frac{1}{2}(1-s)(1 - 2\epsilon)^{L},
\end{equation}
the form of which is a consequence of repeated, independent applications of the Y-flip channel, \eqref{eq:yflip channel}, along with an initial state-prep error, $s$.
The results for 10,000 shots per sequence length are shown in Fig.~\ref{fig:RyPi4RamseyPlot}, and yield fit parameters $s= 3^{+5}_{-2} \times 10^{-5}$ and $\epsilon = 1.1^{+4}_{-4} \times 10^{-4}$, corresponding to a logical $R_y(-\pi/4)$ infidelity of $7^{+3}_{-2}\times10^{-5}.$ 

There is some possible concavity in the data, although the standard deviations of the survival rates make this observation far from conclusive, which suggests either an uneven application of memory error per rotation due to program compilation, or the presence of errors which persist and accumulate across the effective QED cycles that are performed after each rotation, without being converted to logical $\bar{Y}$ errors. One candidate for such errors would be data qubit leakage during two qubit gates, which could be a significant source of error at long sequence lengths. Our magic state protocol is not fault-tolerant to this kind of error as we elect not to perform our leakage detection gadget~\cite{Moses2023} in these experiments, as it adds noise which is significant relative to the fidelity of the magic states and $R_y(\pi/4)$ rotations derived from our protocol. Although this would not address leakage during the magic state protocol itself, we remark that the $R_y(\pi/4)$ rotation could be made fault-tolerant to leakage by designing a gate teleportation circuit which switches the data qubits and magic state qubits while performing the gate. For an example of such a circuit, see~\cite{mayer2024}. 

For each $R_y(\pi/4)$ gate, a single attempt is made to prepare a magic state. If this fails, we abort the protocol. On a device with enough qubits for many magic state factories and a dynamic compiler, we could increase the acceptance rate of the $R_y(\pi/4)$ gate by employing a repeat-until-success strategy and increasing the number of attempts to prepare a magic state per gate. We can approximate what the acceptance rate of the gate teleportation would be in this case by looking at just the post-selection rate of the gate teleportation checks themselves. This is the probability of a non-trivial syndrome occurring in either the $Y$ measurement on the top qubit, the $Z$ measurement on the bottom qubit, or the round of $X$ syndrome extraction on the bottom qubit in the gate teleportation circuit (the dashed region in Fig.~\ref{fig:magicprepcombined}c). The survival and acceptance rates for each sequence length are shown in Table \ref{table: Rypi4} in the Additional Figures section. 

These experiments were performed on H1-1, which has twenty qubits, six of which were initialized via the circuit in Fig. \ref{fig:FTStatePrepH6.pdf}, and twelve of which were repeatedly reset to prepare two logical magic states in parallel. The remaining two qubits were ancilla which were repeatedly reset to perform all checks and syndrome extraction.

\subsection{Benchmarking the Controlled-Hadamard}

\begin{figure}
    \centering
    \includegraphics[width=0.8\linewidth]{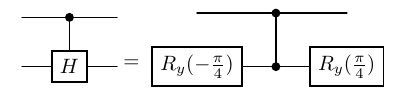}
    \caption{Decomposition of the controlled-Hadamard gate into $R_y(\pm \pi/4)$ rotations which can be implemented using magic states and a $CZ$ gate which is transversal in the [[6,2,2]] code.}
    \label{fig:CHEquality}
\end{figure}

In addition to using these magic states to perform non-Clifford $R_y(\pm \pi/4)$ rotations, we also can use them to fault-tolerantly perform a logical controlled-Hadamard (CH) gate between two qubits in different [[6,2,2]] code blocks. The CH-gate is a non-Clifford gate, and can be performed using $R_y(\pm \pi/4)$ rotations implemented via magic states generated by the experimental protocol and a transversal controlled-Z between the two code blocks (Fig.~\ref{fig:CHEquality}). 

In this section, we describe benchmarking a fault-tolerant logical CH-gate implemented on a single logical qubit in each [[6,2,2]] code block and compare this to a physical CH-gate on H1-1. We show that the infidelity of the logical gate is well below the infidelity of the physical two-qubit gate. Since this logical controlled-Hadamard is the main gate used in the next level of the protocol, this experiment also shows that we are well below threshold for the protocol. To obtain bounds on the physical and logical fidelity, we evaluated the CH gate on 12 carefully chosen input states. The states were chosen to ideally have deterministic measurement outcomes to minimize the number of shots needed to verify logical infidelities on the order of $10^{-4}$. The set of states, $\mathcal{S}$, consists of all combinations of states where the control qubit is in the logical $Z$ basis and the target qubit in the $X$ or $Z$ basis, along with all combinations where the control qubit is in the $X$ basis and the target qubit is in the $H$ basis.
All logical input states are prepared fault-tolerantly either using the circuit in Fig \ref{fig:FTStatePrepH6.pdf} or the experimental magic state protocol. For benchmarking the $\ket{\pm, H^\pm}_L$ states, a final fault-tolerant $R_y(\pi/4)$ rotation is performed on the second qubit before measuring out. In the logical experiment, out of the twenty qubits available on H1, twelve were used to encode the inputs, six were used to encode the magic states and repeatedly reset, and the remaining two were used as ancillae. 10,000 shots were taken per input at two different times and physical and logical submissions/different inputs were interleaved. For the physical jobs, 10,000 shots per input were taken as well and all twenty qubits were gated by grouping the qubits into pairs and applying a physical CH gate in each shot. Additionally, these CH jobs were interleaved with jobs that repeatedly measured the $\ket{0}$ and $\ket{1}$ state so that we could factor out measurement error from the physical jobs. For the physical experiment, all states were rotated to the $\ket{0}$ state after application of the $CH$-gate because measurement error is less for the $\ket{0}$ state than the $\ket{1}$ state~\cite{Olmschenk2007} (in our experiments, we found measurement errors of $1.32^{+3}_{-4}\times 10^{-3}$ for the $\ket{0}$ state $3.66^{+5}_{-6}\times 10^{-3}$ for the $\ket{1}$ state).  The results are summarized in Table \ref{table: CH}. These state infidelities impose convex constraints on the space of Choi matrices for the noisy CH-gate processes, and we solve a semidefinite programming problem to obtain the minimum and maximum average fidelities for each, taking into account measurement error in the physical case (see Appendix A of \cite{ryananderson2022} for more details). We obtain that the infidelity of the logical CH-gate is at most $2.3^{+9}_{-9}\times 10^{-4}$ and that the infidelity of the physical CH-gate is at least $1.0^{+1}_{-1}\times 10^{-3}$, demonstrating a gate fidelity that is beyond the break-even point for a fault-tolerant non-Clifford gate.

\begin{table*}
\renewcommand{\arraystretch}{1.4}
\setlength\tabcolsep{4pt}
\centering
\begin{tabular}{ |c|c|c|c|c|c|c| } 
 \hline
 \multirow{2}{*}{\bf Input State} & \multicolumn{2}{|c|}{\bf CH Infidelity} & \multicolumn{2}{|c|}{\bf Errors per $10^4$ shots} & \multicolumn{1}{|c|}{\bf Shots Accepted} &{\bf Gate Accept Rate}  \\
 \cline{2-7}
 & Physical & Logical & Physical & Logical & Logical & Logical \\
 \hline
 $\ket{0,0}$ & $(3.7^{+2}_{-2})\times 10^{-3}$ & $(3^{+2}_{-2})\times 10^{-4}$ & 36.9 & 2 & 7624  & $85.3_{-4}^{+3}\%$\\ 
 \hline
 $\ket{0,1}$ & $(4.1_{+2}^{+2})\times 10^{-3}$ & $(0^{+1}_{-0})\times 10^{-4}$ &  41.1 & 0 & 7830 & $86.0_{-3}^{+4}\%$\\ 
 \hline
 $\ket{1,0}$ & $(3.1^{+2}_{-2})\times 10^{-3}$ & $(1.3^{+20}_{-8})\times 10^{-4}$ & 31.1 & 1 & 7831 & $85.6_{-4}^{+4}\%$\\ 
 \hline
 $\ket{1,1}$ & $(5.3^{+2}_{-2})\times 10^{-3}$ & $(1.3^{+21}_{-8})\times 10^{-4}$ & 35.2 & 1 & 7736 & $85.7_{-4}^{+4}\%$\\ 
 \hline
  $\ket{+,H^+}$ & $(4.0^{+2}_{-2})\times 10^{-3}$ & $(0^{+1}_{-0})\times 10^{-4}$ & 39.7 &  0 & 6679 & $87.0_{-4}^{+4}\%$\\ 
 \hline
  $\ket{+,H^-}$ & $(4.2^{+2}_{-2})\times 10^{-3}$ & $(4_{-1}^{+4})\times 10^{-4}$ & 42.4 & 3 & 6732 & $87.4_{-4}^{+4}\%$\\ 
 \hline
  $\ket{-,H^+}$ & $(3.5^{+2}_{-2})\times 10^{-3}$ & $(0^{+1}_{-0})\times 10^{-4}$ & 35.2 & 0 & 6881 & $87.8_{-4}^{+3}\%$\\ 
 \hline
 $\ket{-,H^-}$ & $(3.8^{+2}_{-2})\times 10^{-3}$ & $(3_{-2}^{+3})\times 10^{-4}$ & 38.0 & 2 & 6719 & $87.1_{-3}^{+4}\%$\\ 
 \hline
  $\ket{0,+}$ & $(3.7^{+2}_{-2})\times 10^{-3}$ & $(0^{+1}_{-0})\times 10^{-4}$ & 37.4 & 0 & 7479 & $83.6_{-4}^{+4}\%$\\ 
 \hline
 $\ket{0,-}$ & $(4.0^{+2}_{-2})\times 10^{-3}$ & $(0^{+1}_{-0})\times 10^{-4}$ & 39.7 & 0 & 7298 & $81.1_{-4}^{+5}\%$\\ 
 \hline
  $\ket{1,+}$ & $(3.3^{+2}_{-2})\times 10^{-3}$ & $(4_{-2}^{+3})\times 10^{-4}$ & 32.9 & 3 & 7786 & $84.2_{-4}^{+4}\%$\\ 
 \hline
 $\ket{1,-}$ & $(3.9^{+2}_{-2})\times 10^{-3}$ & $(1.3^{+21}_{-8})\times 10^{-4}$ & 38.5 & 1 & 7606 & $84.0_{-4}^{+4}\%$\\ 
 \hline
\end{tabular}

\caption{{\bf CH Infidelity} State infidelities from physical and logical controlled-Hadamard on H1-1 for different input states. For both physical and logical experiments, 10,000 shots were submitted per state. In the physical experiments, all 20 qubits in the quantum processor were gated leading to an overall 100,000 physical CH gates for each input state. In the logical experiments, shots were accepted if no errors were detected during the controlled-Hadamard gate as well as if the input states and magic states were prepared successfully. 
Under ``Gate Accept Rate", we separately report the percentage of shots in which no errors are detected during the CH gate, conditioned on successful preparation of the required resource $\ket{H^+}$ states (since these can be prepared offline using a repeat-until-success protocol). 
Benchmarking the $\ket{\pm,H^{\pm}}_L$ states requires an additional $R_y(\pi/4)_L$ rotation on the target qubit after the CH gate. For these states, the reported gate acceptance rate is also conditioned on the success of this additional rotation. \label{table: CH}
}
\end{table*}

\section{Self-Concatenation}
\begin{figure}
    \includegraphics[width=0.5\textwidth]{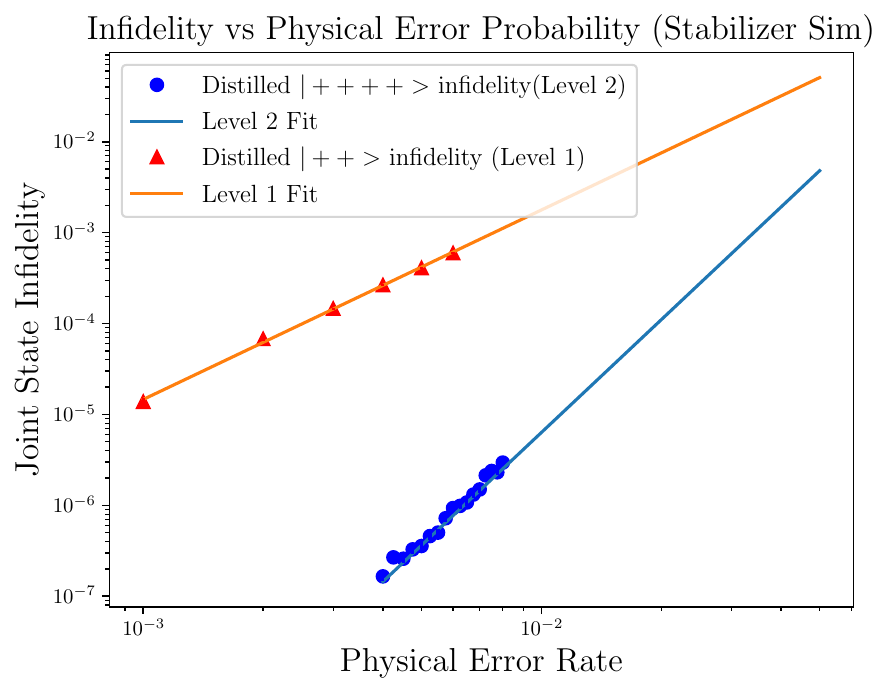}
    \caption{The failure probabilities for the $\ket{+}$ distillation protocol and its concatenation in circuit-level stabilizer simulations. A log plot shows the expected scaling rates of $O(p^2)$ and $O(p^4)$ respectively. Using linear regression, we obtain the estimates $26^{+3}_{-3} \times p^{2.08^{+3}_{-3}}$ and $1000^{+500}_{-500} \times p^{4.1^{+1}_{-1}}$ for the scaling of the logical error of the level 1 and level 2 protocols, respectively.}
    \label{fig:Stabilizer Simulation Plots}
\end{figure}

The experimental results above have been presented for a distance-2 quantum error \emph{detecting} code. However, this protocol may be self-concatenated to produce four magic states in a [[36,4,4]] quantum error \emph{correcting} code by replacing the physical operations in the unconcatenated level-1 protocol shown in Fig.~\ref{fig:magicprepcombined}c with transversal logical operations and the $\ket{H}, \ket{0},$ and $\ket{+}$ input states with $\ket{H^+,H^+}_L, \ket{0,0}_L$ and $\ket{+,+}_L$ states, which are fault-tolerantly prepared either using the level-1 protocol  or the state-prep circuit in Fig. \ref{fig:FTStatePrepH6.pdf}. The logical controlled-H gates are decomposed as in Fig. \ref{fig:CHEquality}, where the CZ gate is transversal and the $R_y(\pm \pi/4)$ rotations are performed via level-1 magic states.  Since the number of qubits in the concatenated protocol exceeds the size of existing Quantinuum quantum processors, in lieu of an experimental implementation, we estimate its fidelity through simulations. Since simulations of large non-Clifford circuits are challenging, we instead employ the  trick in ~\cite{Goto_magic} and perform stabilizer simulations of the above protocol, replacing the $\ket{H^+,H^+}_L$ state to be distilled with the $\ket{+,+}_L$ state as well as the non-Clifford controlled-Hadamard gate with CNOT gates. This will allow us to efficiently simulate the protocol concatenated with itself, which we expect to have failure rate $O(p^4)$ since, informally, either both input magic states would have to fail with probability $O(p^2)$ each or two of the logical two qubit gates in the protocol would have to fail, again with probability $O(p^2).$ To test this scaling relation, we perform simulations with a parameterized noise model in which the ratio of different error rates is comparable to those of existing trapped-ion hardware, but with an overall tunable scale: denoting the two qubit gate error parameter as $p$, we take measurement error $p$, memory error modeled by depolarizing noise with strength $p/5$, and neglect the errors due to single-qubit gates and state preparation, since these are $\sim$100 times lower in H1-1 than other sources of error~\cite{h1specs}. 


For the concatenated protocol, we implement the CNOT gates by using two lower-level distilled $\ket{+}_L$ states to implement $R_Y(\pi/2)$ rotations around a transversal CNOT, analogous to the controlled-Hadamard implementation in Fig.~\ref{fig:CHEquality} so that the stabilizer version of the protocol has the same structure as the original. The results of the stabilizer simulation are plotted in Fig.~\ref{fig:Stabilizer Simulation Plots}. For the concatenated protocol, we have a predicted failure rate of $4.8\times10^{-10}$ at a gate error rate of $p=10^{-3}$ with an acceptance rate of $67\%$ and $3.6\times 10^{-14}$ at an error rate of $p=10^{-4}$ with an acceptance rate of $96\%$. This level-2 protocol uses 160 qubits total and produces four distilled states, yielding a footprint of 40 physical qubits per magic state. The one significant difference between the $\ket{+}$ protocol and the $\ket{H}$ protocol is that corrections for the $R_y(\pi/2)$ rotations were applied by updating the Pauli Frame so that Stim circuit sampling could be used, which is orders of magnitude faster than Stim's tableau simulator, which would need to be used for feed-forward measurement. We found this speed necessary to take an order of 1 billion samples per point in the level-2 concatenation plot. 

Finally, we remark that the error rates for distilling the $\ket{H}_L$ may be higher than those for distilling the $\ket{+}_L$ state in this section, see Assumption 3.1 in \cite{gidney2024}, which suggests an increase by a factor of two, although it is unclear whether this assumption holds at very low physical error rates or for a fault-distance of $4$ as in this work. For this reason, fast, non-Clifford simulators will be important in developing this protocol.

In the case of the $\ket{H}$ protocol, one will have to perform partial $R_y(\pi/2)$ corrections $50\%$ of the time, which could be implemented using $\ket{+0}_L$ ancilla which would require an expected cost of $9$ additional physical qubits per magic state. However, in our simulations of the concatenated protocol, no attempt was made to reuse qubits, so this could presumably offset some of this cost.

\section{Discussion}
We have proposed a fault-tolerant magic state protocol that produces two magic states in a [[6,2,2]] code using eight physical qubits. Using a modification of this protocol, we are able to experimentally produce and validate high-fidelity magic states using a trapped-ion, QCCD architecture. We break even on a non-Clifford two qubit gate, the controlled-Hadamard, achieving a substantially lower infidelity ($2.3^{+9}_{-9}\times10^{-4}$) on the logical level than in the physical controlled-Hadamard $(1.0^{+1}_{-1}\times10^{-3}$). Since the logical controlled-Hadamard is the key non-transversal component in the concatenated protocol, we thus show that we are below threshold for this protocol. 

When this protocol is concatenated with itself, it produces 4 magic states in a $[[36,4,4]]$ code block. Circuit-level simulations indicate that these magic states are expected to have infidelities of order $10^{-10}$ to $10^{-14}$ depending on the physical two-qubit gate infidelity. This concatenated protocol requires approximately 40 physical qubits per magic state and runs in depth 15, yielding a magic state factory with a small spacetime footprint. 

However, in assessing the overall cost of the factory, we should also include an ``escape stage" where the magic states are injected into a higher-distance code which will be used for computation. For magic state cultivation, this step is the most expensive one due to the low encoding rate of the surface code. For instance, escaping from a distance of 5 to 15 in the surface code would require at least 200 additional data qubits and 224 ancilla qubits for syndrome extraction, as well as a circuit of non-trivial depth due to multiple rounds of syndrome extraction.

After preparing four magic states in $[[36,4,4]]$, before escaping to a higher-distance code, we first propose code switching from $[[36,4,4]]$ to $[[16,4,4]]$ to reduce the encoding overhead. This can be implemented via a fault-tolerant code switching strategy between $[[6,2,2]]$ and $[[4,2,2]]$ introduced in Appendix~\ref{app: code switch}. Next, four such $[[16,4,4]]$ codeblocks, each containing 4 magic states, can be teleported into a $[[256,16,16]] = [[4^4, 2^4, 2^4]]$ code via level-raising teleportation, whereby distance 16 Bell states are used to teleport the magic states into the higher distance codeblock. The entire protocol generates 16 magic states using 127 qubits per magic state and can be done in depth 5, assuming that two ancillary $[[256,16,16]]$ code blocks are prepared in parallel with the magic state protocol. The qubit overhead is dominated by the 1015 physical qubits required to fault-tolerantly initialize two $[[256,16,16]]$ code blocks by adapting the methods in \cite{Goto_2024}. This qubit overhead is considerably less than that required for comparable-distance magic state cultivation protocols for surface codes~\cite{gidney2024}, and can significantly reduce the overheads required for large-scale fault-tolerant computation.

The experimental validation of the first concatenation level of this protocol lays the groundwork for performing scalable, universal computations with non-Clifford gates in a family of high-distance codes with multiple logicals.

 \noindent\textbf{Acknowledgments:} We thank the entire Quantinuum team for their many contributions that made this research possible.

\noindent\textbf{Author Contributions:} 
S.D. devised the magic state preparation protocols and circuits, conducted numerical simulations and experiments. All authors contributed to results discussion, data analysis, and manuscript preparation.

\bibliography{refs}{}

\newcommand{\etalchar}[1]{$^{#1}$}
\begin{thebibliography}{MRAB{\etalchar{+}}24}

\bibitem[Ama19]{amaro2019}
David Amaro.
\newblock Stabgraph.
\newblock \url{https://github.com/davamaro/stabgraph},, July 2019.

\bibitem[BH12]{Bravyi2012}
Sergey Bravyi and Jeongwan Haah.
\newblock Magic-state distillation with low overhead.
\newblock {\em Phys. Rev. A}, 86:052329, Nov 2012.

\bibitem[BK05]{Bravyi2005}
Sergey Bravyi and Alexei Kitaev.
\newblock Universal quantum computation with ideal clifford gates and noisy ancillas.
\newblock {\em Phys. Rev. A}, 71:022316, Feb 2005.

\bibitem[CN20]{Chamberland_2020}
Christopher Chamberland and Kyungjoo Noh.
\newblock Very low overhead fault-tolerant magic state preparation using redundant ancilla encoding and flag qubits.
\newblock {\em npj Quantum Information}, 6(1), October 2020.

\bibitem[DBKB{\etalchar{+}}25]{Daguerre:2025}
Lucas Daguerre, Robin Blume-Kohout, Natalie Brown, David Hayes, and Isaac Kim.
\newblock {Experimental demonstration of high-fidelity logical magic states from code switching}.
\newblock 2025.
\newblock To appear.

\bibitem[E{\etalchar{+}}21]{Egan:2020kdu}
Laird Egan et~al.
\newblock {Fault-tolerant control of an error-corrected qubit}.
\newblock {\em Nature}, 598:281--286, 2021.

\bibitem[EK09]{Eastin_2009}
Bryan Eastin and Emanuel Knill.
\newblock Restrictions on transversal encoded quantum gate sets.
\newblock {\em Physical Review Letters}, 102(11), March 2009.

\bibitem[FMMC12]{Fowler2012}
Austin~G. Fowler, Matteo Mariantoni, John~M. Martinis, and Andrew~N. Cleland.
\newblock Surface codes: Towards practical large-scale quantum computation.
\newblock {\em Phys. Rev. A}, 86:032324, Sep 2012.

\bibitem[G{\etalchar{+}}24]{Gupta:2023zei}
Riddhi~S. Gupta et~al.
\newblock {Encoding a magic state with beyond break-even fidelity}.
\newblock {\em Nature}, 625(7994):259--263, 2024.

\bibitem[GF19a]{Gidney_2019}
Craig Gidney and Austin~G. Fowler.
\newblock Efficient magic state factories with a catalyzed $\ket{CCZ}$ to 2$\ket{T}$ transformation.
\newblock {\em Quantum}, 3:135, April 2019.

\bibitem[GF19b]{gidney2019flexible}
Craig Gidney and Austin~G. Fowler.
\newblock Flexible layout of surface code computations using autoccz states, 2019.

\bibitem[Got14]{Goto_magic}
Hayato Goto.
\newblock Step-by-step magic state encoding for efficient fault-tolerant quantum computation.
\newblock {\em Scientific reports}, 4:7501, 12 2014.

\bibitem[Got24]{Goto_2024}
Hayato Goto.
\newblock High-performance fault-tolerant quantum computing with many-hypercube codes.
\newblock {\em Science Advances}, 10(36), September 2024.

\bibitem[GSJ24]{gidney2024}
Craig Gidney, Noah Shutty, and Cody Jones.
\newblock Magic state cultivation: growing t states as cheap as cnot gates, 2024.

\bibitem[HIF24]{hirano2024}
Yutaka Hirano, Tomohiro Itogawa, and Keisuke Fujii.
\newblock Leveraging zero-level distillation to generate high-fidelity magic states, 2024.

\bibitem[ITHF24]{itogawa2024efficient}
Tomohiro Itogawa, Yugo Takada, Yutaka Hirano, and Keisuke Fujii.
\newblock Even more efficient magic state distillation by zero-level distillation, 2024.

\bibitem[Jon13]{Jones2013}
Cody Jones.
\newblock Multilevel distillation of magic states for quantum computing.
\newblock {\em Physical Review A}, 87(4), April 2013.

\bibitem[Kni05]{Knill_2005}
E.~Knill.
\newblock Quantum computing with realistically noisy devices.
\newblock {\em Nature}, 434(7029):39–44, March 2005.

\bibitem[KSU24]{Kim:2024vmw}
Younghun Kim, Martin Sevior, and Muhammad Usman.
\newblock {Magic State Injection on IBM Quantum Processors Above the Distillation Threshold}.
\newblock 12 2024.

\bibitem[L{\etalchar{+}}24]{Lacroix:2024vls}
Nathan Lacroix et~al.
\newblock {Scaling and logic in the color code on a superconducting quantum processor}.
\newblock 12 2024.

\bibitem[LTF{\etalchar{+}}25]{lee2025}
Seok-Hyung Lee, Felix Thomsen, Nicholas Fazio, Benjamin~J. Brown, and Stephen~D. Bartlett.
\newblock Low-overhead magic state distillation with color codes, 2025.

\bibitem[LZ22]{leverrier2022}
Anthony Leverrier and Gilles Zémor.
\newblock Quantum tanner codes, 2022.

\bibitem[M{\etalchar{+}}22]{Marques:2021kev}
J.~F. Marques et~al.
\newblock {Logical-qubit operations in an error-detecting surface code}.
\newblock {\em Nature Phys.}, 18(1):80--86, 2022.

\bibitem[MBA{\etalchar{+}}23]{Moses2023}
S.~A. Moses, C.~H. Baldwin, M.~S. Allman, R.~Ancona, L.~Ascarrunz, C.~Barnes, J.~Bartolotta, B.~Bjork, P.~Blanchard, M.~Bohn, J.~G. Bohnet, N.~C. Brown, N.~Q. Burdick, W.~C. Burton, S.~L. Campbell, J.~P. Campora, C.~Carron, J.~Chambers, J.~W. Chan, Y.~H. Chen, A.~Chernoguzov, E.~Chertkov, J.~Colina, J.~P. Curtis, R.~Daniel, M.~DeCross, D.~Deen, C.~Delaney, J.~M. Dreiling, C.~T. Ertsgaard, J.~Esposito, B.~Estey, M.~Fabrikant, C.~Figgatt, C.~Foltz, M.~Foss-Feig, D.~Francois, J.~P. Gaebler, T.~M. Gatterman, C.~N. Gilbreth, J.~Giles, E.~Glynn, A.~Hall, A.~M. Hankin, A.~Hansen, D.~Hayes, B.~Higashi, I.~M. Hoffman, B.~Horning, J.~J. Hout, R.~Jacobs, J.~Johansen, L.~Jones, J.~Karcz, T.~Klein, P.~Lauria, P.~Lee, D.~Liefer, S.~T. Lu, D.~Lucchetti, C.~Lytle, A.~Malm, M.~Matheny, B.~Mathewson, K.~Mayer, D.~B. Miller, M.~Mills, B.~Neyenhuis, L.~Nugent, S.~Olson, J.~Parks, G.~N. Price, Z.~Price, M.~Pugh, A.~Ransford, A.~P. Reed, C.~Roman, M.~Rowe, C.~Ryan-Anderson, S.~Sanders, J.~Sedlacek, P.~Shevchuk, P.~Siegfried,
  T.~Skripka, B.~Spaun, R.~T. Sprenkle, R.~P. Stutz, M.~Swallows, R.~I. Tobey, A.~Tran, T.~Tran, E.~Vogt, C.~Volin, J.~Walker, A.~M. Zolot, and J.~M. Pino.
\newblock A race-track trapped-ion quantum processor.
\newblock {\em Phys. Rev. X}, 13:041052, Dec 2023.

\bibitem[MRAB{\etalchar{+}}24]{mayer2024}
Karl Mayer, Ciarán Ryan-Anderson, Natalie Brown, Elijah Durso-Sabina, Charles~H. Baldwin, David Hayes, Joan~M. Dreiling, Cameron Foltz, John~P. Gaebler, Thomas~M. Gatterman, Justin~A. Gerber, Kevin Gilmore, Dan Gresh, Nathan Hewitt, Chandler~V. Horst, Jacob Johansen, Tanner Mengle, Michael Mills, Steven~A. Moses, Peter~E. Siegfried, Brian Neyenhuis, Juan Pino, and Russell Stutz.
\newblock Benchmarking logical three-qubit quantum fourier transform encoded in the steane code on a trapped-ion quantum computer, 2024.

\bibitem[OYM{\etalchar{+}}07]{Olmschenk2007}
S.~Olmschenk, K.~C. Younge, D.~L. Moehring, D.~N. Matsukevich, P.~Maunz, and C.~Monroe.
\newblock Manipulation and detection of a trapped ${\mathrm{yb}}^{+}$ hyperfine qubit.
\newblock {\em Phys. Rev. A}, 76:052314, Nov 2007.

\bibitem[P{\etalchar{+}}22]{Postler:2021ddz}
Lukas Postler et~al.
\newblock {Demonstration of fault-tolerant universal quantum gate operations}.
\newblock {\em Nature}, 605(7911):675--680, 2022.

\bibitem[PBP{\etalchar{+}}25]{Pogorelov:2024zvv}
Ivan Pogorelov, Friederike Butt, Lukas Postler, Christian~D. Marciniak, Philipp Schindler, Markus M\"uller, and Thomas Monz.
\newblock {Experimental fault-tolerant code switching}.
\newblock {\em Nature Phys.}, 21(2):298--303, 2025.

\bibitem[PK22]{panteleev2022}
Pavel Panteleev and Gleb Kalachev.
\newblock Asymptotically good quantum and locally testable classical ldpc codes, 2022.

\bibitem[Qua25a]{h1specs}
Quantinuum.
\newblock H1-1 spec sheet.
\newblock \url{https://docs.quantinuum.com/systems/data_sheets/Quantinuum%20H1%20Product%20Data%20Sheet.pdf?_gl=1*8lrrv9*_gcl_au*ODIxMjA1NDgyLjE3NDYxMjY2NzU.}, June 2025.

\bibitem[Qua25b]{papergithub}
Quantinuum.
\newblock Magic-h6.
\newblock \url{https://github.com/CQCL/Magic-H6}, July 2025.

\bibitem[R{\etalchar{+}}24]{Rodriguez:2024bhh}
Pedro~Sales Rodriguez et~al.
\newblock {Experimental Demonstration of Logical Magic State Distillation}.
\newblock 12 2024.

\bibitem[RA{\etalchar{+}}21]{Anderson2021}
C.~Ryan-Anderson et~al.
\newblock Realization of real-time fault-tolerant quantum error correction.
\newblock {\em Phys. Rev. X}, 11:041058, Dec 2021.

\bibitem[RABA{\etalchar{+}}22]{ryananderson2022}
C.~Ryan-Anderson, N.~C. Brown, M.~S. Allman, B.~Arkin, G.~Asa-Attuah, C.~Baldwin, J.~Berg, J.~G. Bohnet, S.~Braxton, N.~Burdick, J.~P. Campora, A.~Chernoguzov, J.~Esposito, B.~Evans, D.~Francois, J.~P. Gaebler, T.~M. Gatterman, J.~Gerber, K.~Gilmore, D.~Gresh, A.~Hall, A.~Hankin, J.~Hostetter, D.~Lucchetti, K.~Mayer, J.~Myers, B.~Neyenhuis, J.~Santiago, J.~Sedlacek, T.~Skripka, A.~Slattery, R.~P. Stutz, J.~Tait, R.~Tobey, G.~Vittorini, J.~Walker, and D.~Hayes.
\newblock Implementing fault-tolerant entangling gates on the five-qubit code and the color code, 2022.

\bibitem[TBG17]{Temme_2017}
Kristan Temme, Sergey Bravyi, and Jay~M. Gambetta.
\newblock Error mitigation for short-depth quantum circuits.
\newblock {\em Physical Review Letters}, 119(18), November 2017.

\bibitem[Wil27]{Wilson1927}
Edwin~B. Wilson.
\newblock Probable inference, the law of succession, and statistical inference.
\newblock {\em Journal of the American Statistical Association}, 22(158):209--212, 1927.

\bibitem[WMI{\etalchar{+}}98]{Wineland98}
D.~J. Wineland, C.~Monroe, W.~M. Itano, D.~Leibfried, B.~E. King, and D.~M. Meekhof.
\newblock Experimental issues in coherent quantum-state manipulation of trapped atomic ions.
\newblock {\em Journal of Research of the National Institute of Standards and Technology}, 103:259--328, May-June 1998.

\bibitem[Y{\etalchar{+}}23]{Ye:2023hxg}
Yangsen Ye et~al.
\newblock {Logical Magic State Preparation with Fidelity beyond the Distillation Threshold on a Superconducting Quantum Processor}.
\newblock {\em Phys. Rev. Lett.}, 131(21):210603, 2023.

\end{thebibliography}
\bibliographystyle{alpha}
\clearpage

\section*{Methods}
\subsection{Statistical analysis}
All figures are reported in the paper with confidence intervals of $68\%$, corresponding to one standard deviation. When estimating the confidence intervals for logical failure rates, to account for skew in these binomial distributions due to errors being rare events, we use the Wilson score interval with $z=1$, which yields a confidence interval of $68\%$ \cite{Wilson1927}. To give confidence intervals for physical error rates, as well as for the acceptance rates in all experiments, we estimate the uncertainty using the standard binomial distribution formula. 

\subsection{Curve Fitting}
To account for rare events, the SPAM and gate error parameters in eq \eqref{eq: Ramsey Curve} were obtained using maximum likelihood estimation instead of least squares regression, which assumes normality of the sample distributions. From this, we obtain an infidelity of $(7^{+3}_{-2})\times10^{-5}$ for the logical $R_y(\pi/4)$ rotation and $(3.1^{+38}_{-5}) \times 10^{-5}$ for the logical SPAM error. The plot of the data along with the fitted curve is shown in Fig.~\ref{fig:RyPi4RamseyPlot}. 

The exponents and coefficients in the scaling rates of the magic state infidelities for the stabilizer simulations of the level 1 protocol and its concatenation were estimated by using Scipy's ``linregress" subroutine on the logarithm of the data. The uncertanties came from the standard error of the linear regression of the estimated slopes and y-intercepts respectively.

\subsection{Logical State Prep}
All experiments benchmark logical gates implemented using the protocol which produces $\ket{H^+,Y^-}_L$. The gates are implemented on the first logical qubit of one or two [[6,2,2]] codeblocks--we view the second qubit as being gauge-fixed via a $Y_0Y_2Y_4$ stabilizer whose value we verify for magic states but ignore for computational qubits. Since many of the inputs will be Paui basis states, we devise a fault-tolerant circuit for producing $\ket{00}_L$ and benchmark its fidelity. We can then generate $\ket{+,+}_L$ using a transversal Hadamard.

\begin{figure}[t]
    \centering
    \includegraphics[width=0.8\linewidth]{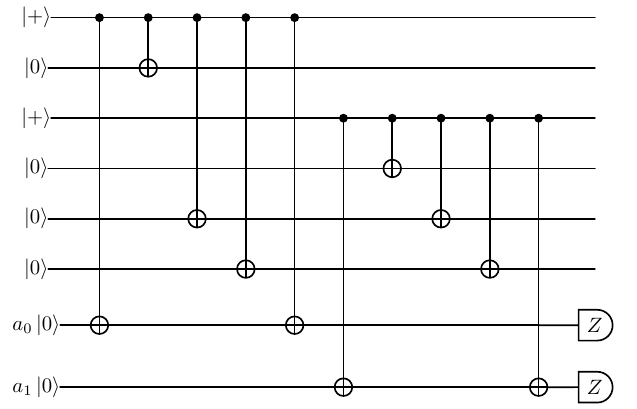}
    \caption{A fault-tolerant state prep circuit for $\ket{00}_L$ in the [[6,2,2]] code. We used StabGraph~\cite{amaro2019} to find a bipartite graph state representation of this stabilizer state. Qubits 0 and 2 are control qubits, and qubits 1,3,4, and 5 are target qubits. We only need to flag qubits 0 and 2 because hook errors on the ``target qubits" are equivalent to weight one errors up to stabilizers and $Z$ logicals.}
    \label{fig:FTStatePrepH6.pdf}
\end{figure}

To find this circuit, we used StabGraph~\cite{amaro2019}, an online package which finds graph state representations for stabilizer states. Crucially, StabGraph can find a bipartite graph state representation for CSS-stabilizer states, allowing us to divide the qubits in the graph states into control and target qubits for CNOTs. This allows us to easily flag the resulting graph state to ensure fault-tolerance. Using this method, we find the fault-tolerant state prep circuit for $\ket{00}_L$ shown in Fig.~\ref{fig:FTStatePrepH6.pdf}. 

\subsection{Data Availability}
The experimental data and processing scripts, as well as the code used to simulate the level-1 and level-2 protocols, are available at \cite{papergithub}.


\begin{table}[t]
\begin{tabular}{ |c|c|c|c| } 
 \hline
  L & Survival Rate & Accept Rate & Accept Rate (Gate)  \\
 \hline
 $0$ & $0.99997^{+2}_{-6}$ & $0.9735^{+9}_{-9}$ & N/A \\ 
 \hline
 $2$ & $0.99986_{-35}^{+5}$ & $0.757^{+4}_{-4}$ & $0.904^{+3}_{-3}$ \\ 
 \hline
 $4$ & $0.9996^{+2}_{-7}$ & $0.549^{+5}_{-5}$ & $0.816^{+5}_{-5}$ \\ 
 \hline
  $6$ & $0.9993^{+3}_{-5}$ & $0.429^{+5}_{-5}$ & $0.742^{+6}_{-6}$ \\ 
 \hline
 $8$ & $0.9987_{-8}^{+5}$ & $0.303^{+5}_{-5}$ & $0.660^{+7}_{-7}$ \\ 
 \hline
\end{tabular}
\caption{The results from the $R_y(-\pi/4)$ Ramsey experiment on H1-1. Using the fault-tolerantly generated magic states, a sequence of $L$ logical $R_y(-\pi/4)$ rotations were performed on the state $\ket{00}_L$ in the [[6,2,2]] code. The survival rate is the percentage of the time $\ket{00}_L$ was sent to the expected Pauli basis state after an even number of rotations. The accept rate is the percentage of shots which are retained over all of the checks in the experiment. This could be increased by repeat-until-success in the future, but is limited by the percentage of shots for which all the gates were performed successfully. This latter percentage is given in the last column.}\label{table: Rypi4}
\end{table}

\clearpage
\appendix 

\section{Details of Magic State Benchmarking \label{app: MSbenchmarkingdetails}}
In this section, we argue that the failure rate of the benchmarking protocol in Fig \ref{fig:magicprepcombined}c is an upper bound for twice the infidelity of the input magic states, scaled by the probability of a non-trivial syndrome, see eq. \eqref{eqn: magic state fidelity bound}. 

In all experiments, we modify the magic state protocol to produce only 1 magic state per code block, without affecting the expected fidelity, by encoding $\ket{H^+,Y^-}_L$ and modifying the $H$-check circuit as in Figure \ref{fig:magicprepcombined}, where $\ket{Y^-} = \frac{1}{\sqrt{2}}(\ket{0} - i\ket{1})$ is the -1 eigenstate of the $Y$ operator. The state being $\ket{H^+,Y^-}_L$ allows us to twirl by applying $\bar{X}_2\bar{H}_1\bar{H}_2$ transversally with probability 1/2 to assume that the noisy magic state is given by an ideal magic state with a logical $\bar{Y}_1$ operator being applied to it with some probability. If we were trying to benchmark the protocol which produced two magic states, $\ket{H^{+},H^{+}}_L$, then we would have to twirl by the non-transversal operators $H\otimes I$ and $I\otimes H$ independently, otherwise off-diagonal terms in the density matrix for the noisy magic states such as $\ket{H^+ H^-}\bra{H^{-},H^{+}}$ will not cancel. These non-transversal operators would add significant noise to the benchmarking procedure relative to the fidelities of the magic states we are trying to measure. 

\begin{figure}[!tb]
    \centering
\includegraphics[width=0.5\linewidth]{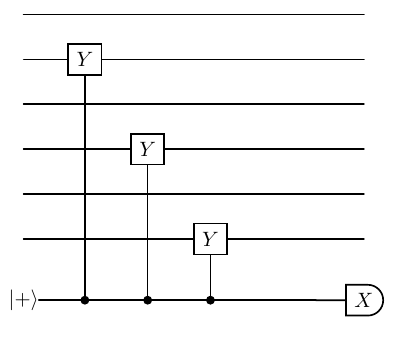}
    \caption{Circuit for measuring the logical $Y_2$ operator in the experimental magic state protocol. This is fault-tolerant because any hook error is a non-logical error that will be detecting in subsequent error detection.}
    \label{fig:H6Y2Check}
\end{figure}

We view the state $\ket{H^+,Y^-}_L$ as being encoded in a gauge-fixed [[6,1,2]] code with an extra stabilizer: $-\bar{Y}_2$ operator. Since the formula we will derive for fidelity has a term which is the probability of a non-zero syndrome, we extract the value of the logical $Y_2$ operator as well, via the circuit in Fig.~\ref{fig:H6Y2Check}.  
This protocol, which encodes $\ket{H^+,Y^-}_L$ using the arbitrary state encoder and then measures the modified $H$-check operator in  Fig.~\ref{fig:magicprepcombined}f and the logical $Y_2$ operator is the magic state protocol we use in all experiments. We expect it to produce magic states of a similar fidelity to the original protocol since it has the same number of two qubit gates arranged in the same pattern. It is fault-tolerant by the same reasoning as for the original magic state protocol.


Before performing the logical rotation, we twirl the magic states by applying the logical $H\otimes X H$ operator with probability $1/2$ (the $X$ operator on the second logical qubit can be regarded as an additional operation to preserve the $Y_1Y_3Y_5$ stabilizer which resulted from fixing the second logical qubit to be $\ket{Y^-}$). 
\begin{widetext}
We write out the component of the density matrix that is in the logical code space of the gauge-fixed $[[6,1,2]]$ code in the $\ket{H},\ket{H^-}$ basis, where $\ket{H^-} = Y\ket{H}$ is the $-1$ eigenvector of the Hadamard operator
\begin{equation}
    (1-p)\ket{H}_L\bra{H}_L + b\ket{H}_L\bra{H^-}_L + c\ket{-H}_L\bra{H}_L + p\ket{-H}_L\bra{-H}_L.
\end{equation}
\end{widetext}

After applying $H \otimes XH$ with probability $\frac{1}{2}$, this will become
\begin{equation}
    (1-p)\ket{H}_L\bra{H}_L + p\ket{-H}_L\bra{-H}_L.
\end{equation}
In other words, the part of $\rho$ in the codespace is either $\ket{H}$ with probability $1-p$ or a logical $Y$ operator has been applied with probability $p$. This probability $p$ is the infidelity of the produced magic state, since, if we were able to probe the produced magic state noiselessly, we would post-select on any non-zero syndromes and these two terms would be the only ones left. Although we are not able to noiselessly probe the state, we nevertheless can use the failure probability of the protocol as an upper bound on the combined infidelity of both magic states.

For both the fault-tolerant and non-fault-tolerant magic states, we assume that they have same noisy density matrix prior to the benchmarking protocol, as both have been produced in the same fashion (the bottom qubit has had the logical $Y_2$ measured, but this is a slight difference and, although the analysis can be modified to take this into account, we assume the infidelities are the same to simplify the argument). After twirling, this is given by

\begin{equation}
    \rho = (1-s)((1-p)\ket{H}_L\bra{H}_L + p\ket{-H}_L\bra{-H}_L) + s\tilde{\rho}
\end{equation}
where $s$ is the probability that the magic state has a non-trivial syndrome and $\tilde{\rho}$ is all of the terms in the density matrix with a component outside of the code space, and $p$ is the infidelity of the magic state as before. We can interpret $p$ as the probability that a logical $Y_1$ has been applied to the magic state. If a $\bar{Y}$ is applied to either magic state at the point of twirling in the benchmarking protocol, then the $Z$ measurement on qubit $2$ will flip and the protocol will fail. The two magic states have not interacted at this point, so a logical $Y$ error on either occurs independently. Therefore, the failure rate of the protocol is at least $2(1-s)(p)(1-p) \sim 2(1-s)(p)$. Denoting by $r$ the probability the magic state protocol fails, the infidelity of each magic state is bounded by:
\begin{equation}\label{eqn: magic state fidelity bound}
    p \leq \frac{r}{2(1-s)}.
\end{equation}
We can estimate the probability of a non-zero syndrome by taking half of the accept rate of the benchmarking protocol. From the fault-tolerant case, where all syndromes are extracted, we obtain $s = .075$.

\section{Fault-tolerantly code switching from [[6,2,2]] to [[4,2,2]] \label{app: code switch}}
In this section, we introduce a procedure for fault-tolerantly switching from the [[6,2,2]] code to the [[4,2,2]] code. We do this because, using the concatenated version of this protocol, we can switch between the $[[36,4,4]]$ code and the $[[16,4,4]]$ code so that we can escape to the family $[[4^L,2^L,2^L]]$, which has a higher encoding rate. 

\begin{figure}[t]
    \centering
    \includegraphics[width=0.5\linewidth]{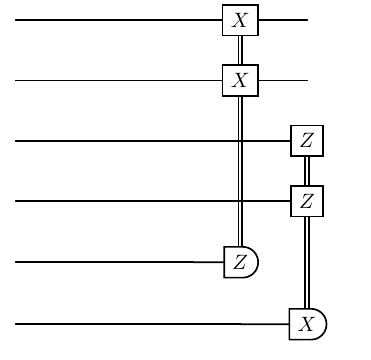}
    \caption{Measuring the fifth qubit of a [[6,2,2]] code in the Z basis and the sixth qubit in the X basis and conditioning logical corrections on these measurements for both of the resulting [[4,2,2]] qubits will non-fault-tolerantly send an arbitrary two qubit state $\ket{\psi_1,\psi_2}_L$ encoded in the [[6,2,2]] code to the same state $\ket{\psi_1,\psi_2}_L$ in the [[4,2,2]] code.}
    \label{fig:nonft_code switch}
\end{figure}

Recall that the stabilizers of the $[[6,2,2]]$ code are generated by $XXXXII, ZZZZII, IIXXXX,$ and $ IIZZZZ$. These contain the stabilizers for the $[[4,2,2]]$ iceberg code on the first 4 qubits. The logical operators for the $[[6,2,2]]$ code are $XIXIXI, IXIXIX, ZIZIZI,$ and $IZIZIZ$ and for the $[[4,2,2]]$ code are $XIIX, XIXI, ZIZI,$ and $IZZI.$ We will now show that the circuit in Fig. \ref{fig:nonft_code switch} will send arbitrary states encoded in the $[[6,2,2]]$ code to ones encoded in the $[[4,2,2]]$ code.

\begin{figure}[t]
    \centering
    \includegraphics[width=1.0\linewidth]{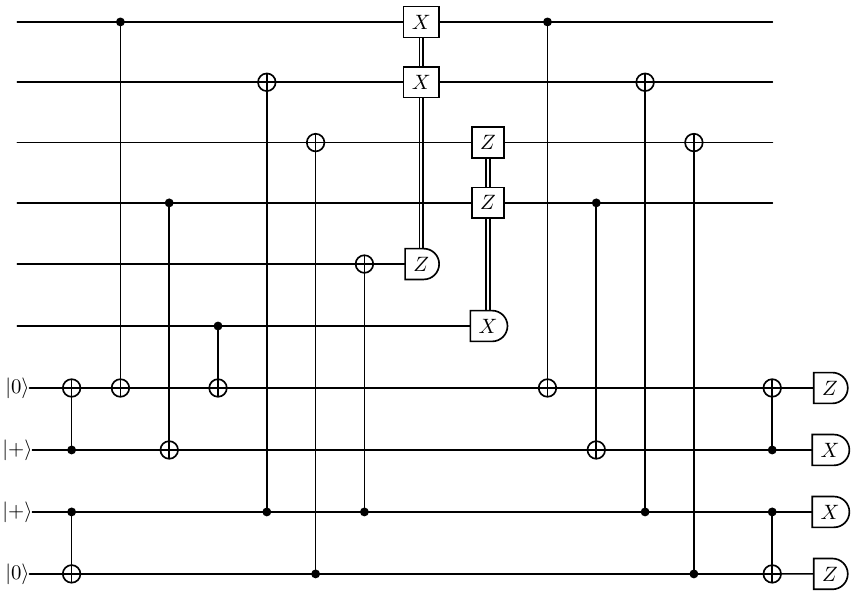}
    \caption{Circuit for making the code switching procedure in Fig. \ref{fig:nonft_code switch}, made fault-tolerant by storing the values of the [[6,2,2]] logical operators and undoing this after measuring the last two qubits of the code and applying the X and Z corrections.}
    \label{fig:ft_code switch}
\end{figure}

To see this, note that if the measurement outcomes for the Z and X measurments are $m_z$ and $m_x$ respectively, we can view this as destroying the $IIZZZZ$ and $IIXXXX$ stabilizers of the [[6,2,2]] code. After these measurements, the stabilizers and logical operators of the $[[6,2,2]]$ code will change as follows:
\begin{align}
\begin{matrix}
    &XXXXII\\
    &ZZZZII\\
    &IIXXXX\\
    &IIZZZZ\\
    \bar{X}_1=&XIXIXI \\
    \bar{X}_2=&IXIXIX\\
    \bar{Z}_1=&ZIZIZI\\
    \bar{Z}_2=&IZIZIZ\\
    \bar{Y}_1=&-YIYIYI\\
    \bar{Y}_2=&-IYIYIY
\end{matrix}
\mapsto
\begin{matrix}
    XXXXII\\
    ZZZZII\\
    (-1)^{m_z}IIIIZI\\
    (-1)^{m_x}IIIIIX\\
    XIIXIX \\
    IXIXIX\\
    ZIZIZI\\
    IZZIZI\\
    YIZXZX\\
    IYZXZX
\end{matrix},    
\end{align}
where we have changed the representation of the logical operators by stabilizers so that they commute with the measurements. We can now use the $(-1)^{m_z}IIIIZI$ and $(-1)^{m_x}IIIIIX$ stabilizers to cancel out the $Z$ and $X$ terms in the last two qubits of the logical operators, leaving us with exactly the logical operators of the $[[4,2,2]]$ code up to sign. The signs will all be $1$ after the conditional $X$ and $Z$ corrections.

As written, this procedure is not fault-tolerant because measurement errors in the Z and X measurements will lead to logical operators being applied. However, such measurement errors can be detected by performing a flagged CNOT controlled by the first logical qubit of the code, targeting a physical ancilla qubit, and repeating this flagging procedure after applying the corrections. If the $X$ corrections are mistakenly applied due to a measurement error, the ancilla will have a measurement value of 1 and we post-select. Otherwise, it will have a measurement value of 0. The same can be done to check if the $Z$ corrections were applied correctly by doing a flagged CNOT from an ancilla to the second logical qubit of the code before and after code switching. This fault-tolerant implementation is shown in Fig. \ref{fig:ft_code switch} and simulated in \cite{papergithub}.


\end{document}